\documentclass[aps,prl,twocolumn,superscriptaddress]{revtex4-1}
\usepackage{amsmath}
\usepackage{graphicx}
\usepackage{braket}
\usepackage{epstopdf}

\graphicspath{{figures/}{}}
\begin{document}

\title{Cavity quantum electrodynamics using a near-resonance two-level system: emergence of the Glauber state}


\author{B. Sarabi}
\affiliation{Laboratory for Physical Sciences, College Park, MD 20740, USA}
\affiliation{Department of Physics, University of Maryland, College Park, MD 20742, USA}
\author{A. N. Ramanayaka}
\affiliation{Laboratory for Physical Sciences, College Park, MD 20740, USA}
\affiliation{Department of Physics, University of Maryland, College Park, MD 20742, USA}
\author{A. L. Burin}
\affiliation{Department of Chemistry, Tulane University, New Orleans, LA 70118, USA}
\author{F. C. Wellstood}
\affiliation{Department of Physics, University of Maryland, College Park, MD 20742, USA}
\affiliation{Joint Quantum Institute, University of Maryland, College Park, MD 20742, USA}
\author{K. D. Osborn}
\affiliation{Laboratory for Physical Sciences, College Park, MD 20740, USA}
\affiliation{Joint Quantum Institute, University of Maryland, College Park, MD 20742, USA}


\date{\today}

\begin{abstract}
Random tunneling two-level systems (TLSs) in dielectrics have been of interest recently because they adversely affect the performance of superconducting qubits. The coupling of TLSs to qubits has allowed individual TLS characterization, which has previously been limited to TLSs within (thin) Josephson tunneling barriers made from aluminum oxide. Here we report on the measurement of an individual TLS within the capacitor of a lumped-element LC microwave resonator, which forms a cavity quantum electrodynamics (CQED) system and allows for individual TLS characterization in a different structure and material than demonstrated with qubits. Due to the reduced volume of the dielectric (80 $\mu$m$^{3}$), even with a moderate dielectric thickness (250 nm), we achieve the strong coupling regime as evidenced by the vacuum Rabi splitting observed in the cavity spectrum. A TLS with a coherence time of 3.2 $\mu$s was observed in a film of silicon nitride as analyzed with a Jaynes-Cummings spectral model, which is larger than seen from superconducting qubits. As the drive power is increased, we observe an unusual but explicable set of continuous and discrete crossovers from the vacuum Rabi split transitions to the Glauber (coherent) state.
\end{abstract}


\maketitle

Cavity quantum electrodynamics (CQED) phenomena, including vacuum Rabi splitting (VRS) \cite{PhysRevLett.51.1925.3} and enhanced spontaneous emission \cite{PhysRevLett.50.1903}, have greatly advanced the understanding of photons coupled to atoms \cite{PhysRevLett.51.1175}, ions \cite{PhysRevLett.74.4091,RevModPhys.75.281} and superconducting qubits \cite{wallraff2004strong,PhysRevA.69.062320,Reed2010}. While the performance of the latter is often limited by random tunneling two-level systems (TLSs) \cite{PhysRevLett.95.210503,PhysRevLett.93.077003,PhysRevLett.93.180401}, these low-energy excitations have also served as local quantum memories \cite{neeley2008process}. In these studies, measurement of individual TLSs properties, including their coherence times \cite{PhysRevB.78.144506,PhysRevLett.105.177001,PhysRevLett.105.230504}, have previously been limited to thin ($\sim1$ nm) layers of aluminum oxide, the prevalent material for Josephson junction tunneling barriers. However, other structures, including capacitors, are used in qubits, and other materials are known to be scientifically interesting due to unconventional TLS properties, {\it e.g.} low TLS density \cite{PhysRevLett.110.135901,5418350,:/content/aip/journal/apl/92/11/10.1063/1.2898887}. It is therefore desirable to characterize {\em individual} TLSs in {\em insulating} structures and materials, without being limited to tunneling-barrier structures.

Here we report on a CQED study with TLSs where the TLSs are coupled to a cavity, which allows us to characterize an individual TLS in an {\em insulating}-thickness film of 250 nm. In our experiment, the cavity is a circuit resonator made from a capacitor containing amorphous silicon nitride dielectric and an inductor. Similar to some amorphous silicon, the type of silicon nitride has a lower density of TLSs when compared to other amorphous solids \cite{PhysRevLett.110.135901,5418350}. By using microscopic volumes of this material, we reach the CQED strong-coupling regime using a single strongly-coupled TLS, and observe VRS below a single photon (on average) in the cavity. We also observe a quantum-to-Glauber (coherent) crossover as the drive power is increased, which results in a wishbone-shaped transmission. This results from two different phenomena emerging from the VRS transitions as the coherent drive power is increased. The weakly-coupled TLSs are also studied, and can be clearly distinguished from the strongly-coupled TLS.

\begin{figure}

\includegraphics[width=0.483\textwidth]{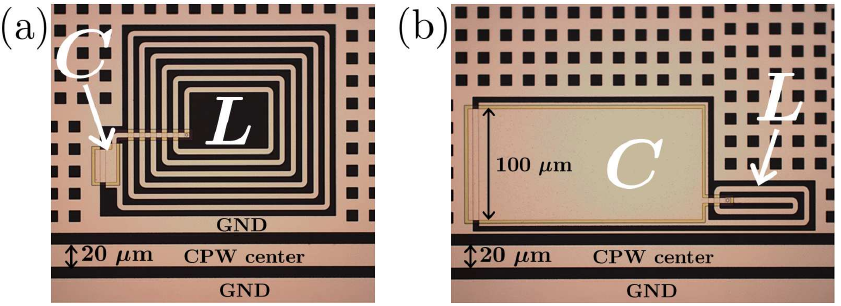}
\caption{Optical image of capacitor $C$ and inductor $L$ for resonators with (a) the smallest (80 $\mu$m$^{3}$) and (b) the largest (5000 $\mu$m$^{3}$) dielectric volumes. Aluminum appears light and the sapphire substrate appears black.}
\label{optical image}
\end{figure}

Microwave resonators (each containing an inductor and a capacitor) were made with superconductor-insulator-superconductor trilayer capacitors having dielectric volumes $V$ ranging from 80 to 5000 $\mu$m$^{3}$ (see Fig. \ref{optical image}). Despite having substantially different volumes, the cavity (lumped-element resonator) frequencies $\omega_{c}/2\pi$ were kept in the 4.6 to 6.9 GHz range. The devices were fabricated from superconducting aluminum films with a 250 nm thick film of amorphous hydrogenated silicon nitride (a-SiN$_{\mathrm{x}}$:H) forming the capacitor dielectric \cite{5418350}. Five resonators were fabricated on a chip and coupled (both inductively and capacitively) to a 20 $\mu$m wide transmission line resulting in a multi-band bandstop transmission.

Each resonator's transmission $S _{21}$ was measured at 25 mK in a dilution refrigerator with a coherent input. The cavity photon number $\bar{n}_{\mathrm{max}}$ changed from approximately $10^{-4}$ to $10^{3}$, where $\bar{n}_{\mathrm{max}}$ is defined as the maximum time-averaged value from a frequency scan at a fixed input power. For the two resonators with the largest insulator volumes, 5000 and 2500 $\mu$m$^{3}$, a standard analysis \cite{:/content/aip/journal/jap/111/5/10.1063/1.3692073} yielded a low-power ($\bar{n}_{\mathrm{max}}\ll1$) loss tangent of $\tan\delta_{0}\equiv\kappa_{\perp}/\omega_{c}\simeq1\times10^{-4}$ where $\kappa_{\perp}$ represents the photon decay rate from internal loss mechanisms set by the weakly-coupled TLSs. At this low-temperature limit, the loss tangent depends on the electric field amplitude $E$ in the dielectric approximately as $\tan\delta=\tan\delta_{0}/\sqrt{1+(E/E_{c})^{2}}$. This follows from the standard model of TLSs \cite{Anderson, Phillips} with excitation energy $\mathcal{E}=\sqrt{\Delta ^{2}+\Delta _{0} ^{2}}$ and standard TLS distribution $d^3N=(P_{0}/\Delta_{0})d\Delta d\Delta _{0}dV$ where $N$ is the TLS number, $P_{0}=3\epsilon_{0}\epsilon_{r}\tan\delta_{0}/\pi p^{2}$, $\Delta _{0}$ represents the tunneling energy and $\epsilon_{0}\epsilon_{r}$ is the dielectric permittivity.  $\Delta$ denotes the offset energy between the two wells, which is perturbed by the interaction energy $\textbf{p}\cdot\textbf{E}=pE\cos\theta$ of the TLS dipole moment $\textbf{p}$, at an angle $\theta$ with respect to the electric field $\textbf{E}$. We measured $E_{c}=4.6$ V/m for the two largest-volume resonators. The same value was found and expected for both resonators since for large (bulk) samples, $E_{c}$ is only dependent on a characteristic TLS coherence time and Rabi frequency. At the critical field in these resonators, the photon number is greater than 1, which is a consequence of small coupling between the TLSs and cavity relative to the TLSs' decay rate, and allows a classical treatment of the field \cite{VonSchickfus1977144}. However, for the smaller volume (larger CQED coupling) regime discussed below, this no longer holds.

\begin{figure}

\includegraphics[width=0.483\textwidth]{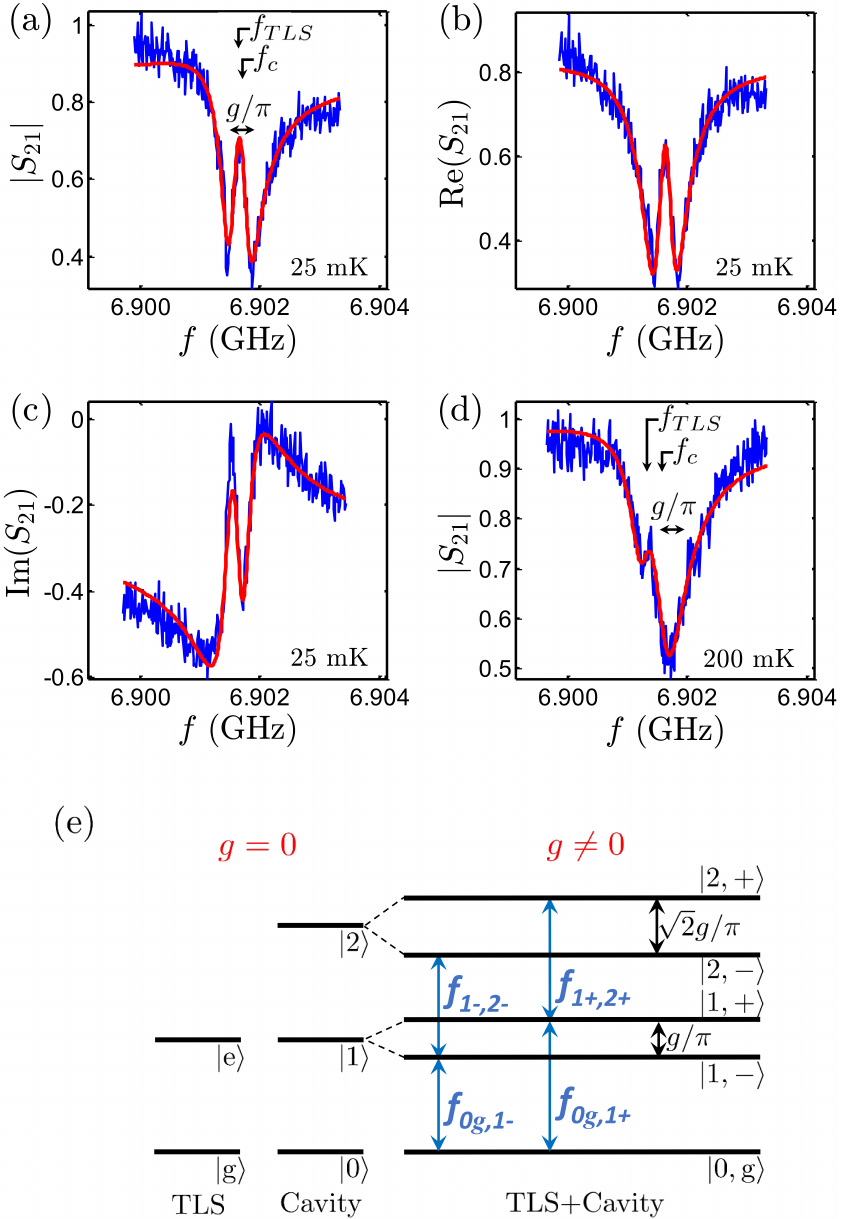}
\caption{(a) Measured $|S_{21}|$ vs. frequency $f$ (blue) and optimum fit (red) at 25 mK. (b) and (c) show corresponding real and imaginary parts of $S_{21}$. (d) Measured (blue) and best fit (red) of $|S_{21}|$ vs. frequency at 200 mK. $f_{c}$ and $f_{\mathrm{TLS}}$ are the cavity and TLS frequencies, respectively. (e) Diagram of the lowest energy levels of the bare TLS and cavity states ($g=0$) on-resonance, and the dressed states ($g\neq0$).}
\label{fitting}
\end{figure}

For the smallest insulator volume, at 80 $\mu$m$^{3}$, at low photon numbers ($\bar{n}_{\mathrm{max}}\ll1$) the transmission magnitude (Fig. \ref{fitting}(a)) showed a second resonant dip, while the real (Fig. \ref{fitting}(b)) and imaginary (Fig. \ref{fitting}(c)) parts revealed that each belongs to a separate resonance loop in the complex plane. This second dip (transition amplitude) is consistent with a single TLS strongly interacting with the cavity, {\it i.e.} CQED. Two intermediate insulator volumes, at 230 and 760 $\mu$m$^{3}$, also showed features consistent with discrete TLSs. Because these devices exhibit CQED effects due to their volumes $V$, we refer to them as micro-$V$ resonators; below we only analyze the 80 $\mu$m$^{3}$ device. Similar to an atom-cavity system, strong-coupling CQED can be achieved when the TLS-resonator coupling $g=\frac{\Delta_{0}}{\mathcal{E}}p\cos\theta\sqrt{\omega_{0}/2\epsilon_{r}\epsilon_{0}\hbar V}$ is similar or greater than $\sqrt{\gamma_{\mathrm{TLS}}(\kappa_{\perp}+\kappa_{\parallel})}$ where $\gamma_{\mathrm{TLS}}$ is the TLS decay rate and $\kappa_{\parallel}$ is the cavity's photon decay rate from coupling to the external transmission line. We now examine how the micro-$V$ resonator can exhibit a single strongly-coupled TLS.

The average number $\bar{N}$ of TLSs in the bandwidth $B$ of the micro-$V$ resonator can be estimated using the standard TLS distribution. TLSs, each with their own $\Delta$ and $\Delta_{0}$, can resonantly exchange excitations with cavity photons at energy $\hbar\omega$. Using a representative angle of $\theta$, we estimate the number of TLSs available for strong coupling as $\bar{N}\simeq2\pi P_{0}V\hbar B$, where $B=\kappa_{\perp}/\pi$ for a critically coupled resonator, and therefore $\bar{N}\propto\tan^{2}\delta_{0}$. Using $\kappa_{\parallel}$ which is near critical coupling ($\kappa_{\parallel}\simeq\kappa_{\perp}$), and the measured value of $p$ from Ref. \cite{Moe3}, we calculate $\bar{N}\simeq1$ observable TLS in the smallest micro-$V$ resonator (given that the corresponding $\gamma_{\mathrm{TLS}}$ satisfies the strong coupling criteria). This is consistent with the micro-$V$ device data taken from multiple cooldowns. We note that this technique could be applied to other materials, {\it e.g.} to achieve $\bar{N}\simeq1$ in air-exposed alumina \cite{:/content/aip/journal/apl/103/16/10.1063/1.4826253} a smaller $V$ can be used to compensate for the larger TLS density.

The Hamiltonian used for further analysis contains the Jaynes-Cummings model \cite{1443594} in the first three terms,
\begin{equation}
\mathcal{H}=\hbar\omega_{c}c^{\dagger}c+i\hbar g(S^{+}c-c^{\dagger}S^{-})+\mathcal{E}S^{z}+\mathcal{H}_{\mathrm{ex}}.
\label{hamiltonian}
\end{equation}
A harmonic oscillator with photon annihilation operator $c$ represents the resonator which is strongly-coupled (with rate $g$) to a TLS with pseudospin operators $S^{-}$, $S^{+}$ and $S^{z}$. The other TLSs are assumed to be weakly-coupled, giving rise to $\kappa_{\perp}$. The external coupling Hamiltonian $\mathcal{H}_{\mathrm{ex}}=\hbar\omega_{d}d^{\dagger}d+\hbar\Omega(d^{\dagger}c+c^{\dagger}d)$ accounts for the coupling to the transmission line with photon annihilation operator $d$.

\begin{figure*}
\includegraphics[width=\textwidth]{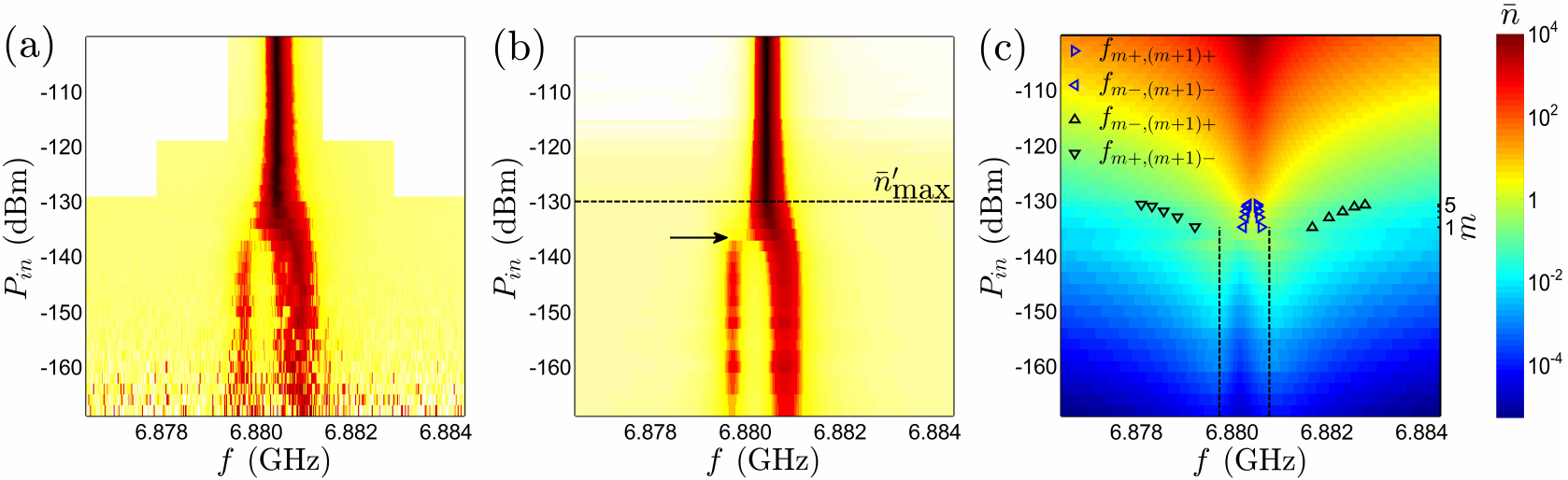}
\caption{(a) False-color plot showing measured transmission $|S_{21}|$ vs. input power $P_{\mathrm{in}}$ and frequency $f$ for the micro-$V$ resonator with $V=80\:\mu$m$^{3}$. (b) Simulated power dependence from theoretical fit to model. $\bar{n}'_{\textrm{max}}\simeq7$ indicates the photon number above which the classical approach is used and the arrow shows the break of $m\geq1$ transitions. (c) False-color plot of the simulated photon occupancy $\bar{n}$ vs. $P_{\mathrm{in}}$ and $f$. The black dashed lines correspond to the vacuum Rabi split transitions at $f_{0\mathrm{g},1-}$ and $f_{0\mathrm{g},1+}$. $f_{m\pm,(m+1)\pm}$ and $f_{m\pm,(m+1)\mp}$ are shown for $m=1$-5.}
\label{power_saturation}
\end{figure*}

At small average photon numbers in the cavity, $\bar{n}_{\mathrm{max}}\ll1$, and in the low-temperature limit ($k_{\mathrm{B}}T\ll\hbar\omega$), the dominant TLS can be treated as an oscillator. At higher temperatures ($k_{\mathrm{B}}T\gtrsim\hbar\omega$) we used a mean field approach and replace $S^{z}$ with its thermodynamic average $\left<S^{z}\right>=-\frac{1}{2}\tanh(\hbar\omega/2k_{\mathrm{B}}T)$. The single-photon transmission, appropriate for $\bar{n}_{\mathrm{max}}\ll1$ is
\begin{equation}
S_{21}\simeq1-\frac{\tilde{\kappa}_{\parallel}/2}{\frac{\tilde{\kappa}_{\parallel}}{2}+\frac{\kappa_{\perp}}{2}+i\left(\omega-\omega_{c}\right)+\frac{g^{2}\tanh(\hbar\omega/2k_{\mathrm{B}}T)}{\frac{\gamma_{\mathrm{TLS}}}{2}+i\left(\omega-\omega_{\mathrm{TLS}}\right)}}.
\label{transmission}
\end{equation}
The complex coupling rate $\tilde{\kappa}_{\parallel}$ has a small imaginary component $\mathrm{Im}(\tilde{\kappa}_{\parallel})\ll\mathrm{Re}(\tilde{\kappa}_{\parallel})(\simeq\kappa_{\parallel})$ and similar to many classical resonators, it plays a negligible role in our device, {\it cf.} Ref. \cite{:/content/aip/journal/jap/111/5/10.1063/1.3692073}.

Under strong coupling conditions, Eq. \ref{transmission} shows a VRS: two distinct transition amplitudes. From Eq. \ref{hamiltonian}, the two vacuum transition frequencies become $2\pi f_{0\mathrm{g},1\pm}=\omega_{c}+\delta/2\pm\sqrt{g^{2}+(\delta/2)^{2}}$, where  $\delta=\omega_{\mathrm{TLS}}-\omega_{c}$ (see Fig. \ref{fitting}(e)). From Eq. \ref{transmission} it follows that a single TLS at resonance with the cavity can be distinguished if its maximum response exceeds the average response of the weakly-coupled TLSs, {\it i.e.} $\chi=\pi P_{0}\hbar V/6\textrm{T}_{1}<1$, where T$_{1}=1/\gamma_{\mathrm{TLS}}$ is the TLS relaxation time.

Figures \ref{fitting}(a)-(c) show a least squares Monte Carlo (LSM) fit to the data using the low-temperature limit of Eq. \ref{transmission} \cite{Bahmanthesis}. The fit yields $f_c=\omega_{c}/2\pi= 6.901689$ GHz and $\kappa_{\perp} = 1.92$ MHz for the resonator, and $f_{\mathrm{TLS}}=\omega_{\mathrm{TLS}}/2\pi = 6.901629$ GHz and T$_{2}=2/\gamma_{\mathrm{TLS}}= 3.2\:\mu$s for the TLS, where the T$_{2}$ is the coherence time of the resonant TLS. This TLS coherence time is at least 3 times larger than previously characterized individual TLSs \cite{PhysRevLett.105.177001,PhysRevLett.105.230504}.

The fit also yields $\kappa_{\parallel}/2\pi=493$ kHz and $g/\pi=366$ kHz, where from $g$ we obtain a transition dipole moment of $p_{\mathrm{min}}=(\Delta_{0}/\mathcal{E})p\cos\theta=1.6$ Debye $=0.34\:e$\AA $\:((\Delta_{0}/\mathcal{E})\cos\theta\leq1)$. This minimum extracted dipole size for the TLS is consistent with a previous measurement of the same material \cite{Moe3}. The spontaneous photon emission rate given by the Purcell effect (with $\kappa_{\parallel}$ and $p_{\mathrm{min}}$) is calculated to be within a factor of 2 of the measured $\gamma_{\mathrm{TLS}}$, indicating that T$_{2}$ may be limited by the photon, rather than phonon, emission.

Figure \ref{fitting}(d) shows data from the same micro-$V$ device at $T=200$ mK. We also fit Eq. \ref{transmission} to this data with $\kappa_{\parallel}$ fixed to the low temperature result. The fit reveals $g/\pi=360$ kHz, showing almost no effect from temperature, while $f_{\mathrm{TLS}} = 6.901318$ GHz and $f_{c} = 6.901576$ GHz show a small shift caused by the weakly-coupled TLS bath, as expected. Unlike the low-temperature result, $\delta$ is now approximately equal to $2g$ and causes an unequal superposition of the bare states.
The high (low) frequency side of the VRS,
$\ket{0,\mathrm{g}}\rightarrow\ket{1,+}$ ($\ket{0,\mathrm{g}}\rightarrow\ket{1,-}$), involves a cavity-like (TLS-like) state and hence the  cavity-measured amplitude at $f_{0\mathrm{g},1+}$ ($f_{0\mathrm{g},1-}$) is larger (smaller) than an equal-superposition state.
Equation \ref{transmission} with the remaining fit values, T$_{1}(\textrm{200 mK})=0.57\:\mu$s and  $\kappa_{\perp}$, allow us to calculate the ratio of the TLS-like transition amplitude on the background of the cavity-like transition amplitude as $4g^2\tanh(\hbar\omega/2 k_{\mathrm{B}}T) $T$_1/\kappa_{\perp}$= 0.67, where $\tanh(\hbar\omega/2k_{\mathrm{B}}T)=0.68$. The T$_{1}(\textrm{200 mK})$ is shorter than expected from phonon emission which scales as $\tanh\left(\hbar\omega/2k_{\mathrm{B}}T\right)$ and predicts T$_{1}(\textrm{200 mK})=1.1\:\mu$s. This discrepancy is qualitatively consistent with additional dephasing expected from spectral diffusion \cite{PhysRevB.16.2879} and the temperature dependence of tunneling barrier TLSs \cite{PhysRevLett.105.230504}, observed here as a larger linewidth ($\sim1/\mathrm{T}_{1}$).

Figure \ref{power_saturation}(a) shows a false-color plot of $|S_{21}|$ (measured at 25 mK) as a function of frequency and input power $P_{\mathrm{in}}$, from a different cooldown of the same micro-$V$ device. For $P_{\mathrm{in}}<-135$ dBm we observe VRS similar to those shown in Fig. \ref{fitting}, as expected from the $m=0\rightarrow1$ (single photon) excitation of the system. At higher powers we drive some other transitions on the Jaynes-Cummings ladder (see Fig. \ref{fitting}(e)), such that $2\pi f_{m\pm,(m+1)\pm}=\omega_{c}\mp\sqrt{g^{2}m+(\delta/2)^{2}}\pm\sqrt{g^{2}(m+1)+(\delta/2)^{2}}$, where the $\mp$ of the second term corresponds to the $m\pm$ state index and the $\pm$ of the third term corresponds to the $(m+1)\pm$ state index.

Similar to the case of Fig. \ref{fitting}(d), in Fig. \ref{power_saturation}(a) we see that the transition at $f_{0\mathrm{g},1+}$ has a larger amplitude than that at $f_{0\mathrm{g},1-}$ because $\ket{1,+}$ has a larger component of $\ket{1,\mathrm{g}}$ than $\ket{0,\mathrm{e}}$. At the highest input power, we must have the Glauber (coherent) state which at frequency $\omega_{c}/2\pi$ is only slightly smaller than $f_{0\mathrm{g},1+}$.  We also notice that as the power is increased there is a continuous crossover from $\ket{0,\mathrm{g}}\rightarrow\ket{1,+}$ to higher-energy transitions eventually reaching the Glauber (coherent) state. These transitions are excited from $\ket{1,+}$ and include the transitions $\ket{m,+}\rightarrow\ket{m+1,+}$ in climbing the Jaynes-Cummings ladder (see $\ket{1,+}\rightarrow\ket{2,+}$ in Fig. \ref{fitting}(e)).

In contrast, as we start from $f_{0\mathrm{g},1-}$, we observe a different behavior which is caused by the detuning ($|f_{c}-f_{0\mathrm{g},1-}|>|f_{c}-f_{0\mathrm{g},1+}|$). A gap of transition amplitude appears between the $\ket{0,\mathrm{g}}\rightarrow\ket{1,-}$ transition and the higher power transitions $\ket{m,-}\rightarrow\ket{m+1,-}$ (the $\ket{m,\pm}\rightarrow\ket{m+1,\pm}$ transitions are included in the (high-power) Glauber state according to the Jaynes-Cummings model). A break between the $\ket{0,\mathrm{g}}\rightarrow\ket{1,-}$ and the higher $\ket{m,-}\rightarrow\ket{m+1,-}$ transitions has been previously observed as a quantum-to-classical crossover in a superconducting qubit-resonator system, but there the crossover to the coherent state, which includes the $\ket{m,+}\rightarrow\ket{m+1,+}$ transitions, was not observed due to the use of an incoherent drive source \cite{PhysRevLett.105.163601}. The break from the $\ket{0,\mathrm{g}}\rightarrow\ket{1,-}$ transition to other transitions has allowed for demonstrations of photon blockade \cite{PhysRevLett.106.243601}.

We analyzed the nonlinear data in high and low input power regimes separately. In both regimes, the frequency scan data was fit at each measurement input power. The low-power regime starts below $P'_{\mathrm{in}}=-130$ dBm where $\bar{n}_{\mathrm{max}}=\bar{n}'_{\mathrm{max}}\simeq7$. For the high power coherent-like state, $f_{m+,(m+1)+}$ is very close to $f_{m-,(m+1)-}$ and the width of the cavity resonance is primarily determined by $\kappa_{\perp}$ (the weakly-coupled TLS bath) which allows a classical field analysis \cite{:/content/aip/journal/jap/111/5/10.1063/1.3692073}. A LSM fit to this regime gave $\omega_{c}/2\pi=6.880434$ GHz, $\kappa_{\parallel}/2\pi=491$ kHz and $\kappa_{\perp}(P_{\mathrm{in}})$, where the weakly-coupled TLS are influenced by $P_{\mathrm{in}}$, similar to previous classical saturation field studies. In the low-power regime, we used a calculation of the density matrix, Eq. \ref{hamiltonian}, using the Lindblad formalism. A LSM fit to this data, using $\kappa_{\parallel}$ and $\omega_{c}$ from above, gave $g/\pi=1.00$ MHz, $\omega_{\mathrm{TLS}}/2\pi=6.880106$ GHz, and T$_{1}=1/\gamma_{\mathrm{TLS}}=325$ ns and the remaining regime for $\kappa_{\perp}(P_{\mathrm{in}})$, which is approximately equal to $\kappa_{\perp}$ of the largest-volume device when $P_{\mathrm{in}}\ll P'_{\mathrm{in}}$.

A combination of the fits and the resulting $\bar{n}$ are shown in Figs. \ref{power_saturation}(b) and \ref{power_saturation}(c), respectively. Notice that $\bar{n}\gtrsim1$ only near $f_{c}$ and at $P_{\mathrm{in}}>-136$ dBm, while at lower powers, $\bar{n}$ has a local maximum in frequency scans at $f_{0\mathrm{g},1-}$ and $f_{0\mathrm{g},1+}$. The transition frequencies $f_{m+,(m+1)+}$ and $f_{m-,(m+1)-}$ are plotted in Fig. \ref{power_saturation}(c) where the vertical placement of $m$ is only suggestive. This shows how the occupancy of the Jaynes-Cummings transitions near $f_{c}$ are populated from $\ket{1,+}$ rather than $\ket{1,-}$ where spontaneous emission can cause the $\ket{2,+}\rightarrow\ket{1,-}$ transition. In this system, the TLS spontaneous emission is small $\gamma_{\mathrm{TLS}}/4(\kappa_{\perp}+\kappa_{\parallel})\simeq0.1\ll1$, and is believed to switch the field phase during $\ket{m,+}\leftrightarrow\ket{m\pm1,-}$ transitions \cite{0954-8998-3-1-003}. The other transitions, $\ket{m,-}\rightarrow\ket{m+1,+}$ and $\ket{m,+}\rightarrow\ket{m+1,-}$, are suppressed due to low occupancy of $\ket{1,\pm}$ ($\bar{n}\ll1$) at the frequency of these transitions, as suggested by the figure.

In conclusion, we have measured and characterized individual TLSs in an insulating film using strong TLS-resonator coupling. The dielectric in our experiment is approximately 2 orders of magnitude thicker than a tunneling barrier used in previous (qubit) studies of individual TLS.

We observe the vacuum Rabi split (VRS) transitions at single-photon powers, and a crossover to the Glauber (coherent) state as the input measurement power is increased. The continuous crossover from one VRS transition, and a discontinuous crossover from the other is explained by a Jaynes Cummings model. This differs from superconducting qubit CQED studies which have an incoherent drive tone to study this crossover, where more than 2 levels of their qubit are used for quantitative comparison to simulations.

The device design allows other dielectrics to be studied in the future, in contrast to studies of TLS in the Josephson tunneling barriers of qubits which (typically) use aluminum oxide. One TLS in our SiN$_{\mathrm{x}}$ film was found to have a coherence time of T$_{2}=3.2$ $\mu$s, which is longer than those in superconducting qubits. The difference may be due to the capacitor structure, which allows the TLS to be isolated from the electrodes, or the material. This T$_{2}$ time is similar to that of the original measurement of the popular transmon qubit and we believe that in the CQED architecture, TLS with longer coherence times will be found. One TLS was found to have strong coupling at $g/\pi=1$ MHz, such that the VRS transitions are well resolved, and in the future one transition (from the CQED system) could be operated as a qubit without the use of a Josephson-junction qubit.

The authors thank C. Lobb, R. Simmonds, B. Palmer, Y. Rosen, M. Stoutimore, M. Khalil and S. Gladchenko for many useful discussions. A. Burin acknowledges support through Army Research Office Grant vv911NF-13-1-0186, the LA Sigma Program, and the NSF EPCORE LINK Program.

\bibliography{references}

\end{document}